# The effect of electric field on multiple exciton generation in lead chalcogenide nanocrystals

*Mahdi Gordi[1], Mohammad Kazem Moravvej-Farshi[1,*]*

[1] Advanced Devices Simulation Lab (ADSL), Faculty of Electrical and Computer Engineering, University of Tarbiat Modares, P. O. Box 14115-194, Tehran 1411713116, Iran.

*Hamidreza Ramezani[2,*]*

[2] Department of Physics and Astronomy, University of Texas Rio Grande Valley, Brownsville, TX 78520, USA.

Corresponding Authors:

*Emails: moravvej@modares.ac.ir and hamidreza.ramezani@utrgv.edu

**ABSTRACT:** Unique properties of lead chalcogenides have enabled multiple exciton generation (MEG) in their nanocrystals that can be beneficial in enhancing the efficiency of the third generation solar cells. Although the intrinsic electric field plays an imperative role in a solar cell, its effect on the multiple exciton generation (MEG) has been overlooked, so far. Using EOM-CCSD as a many-body approach, we show that any electric field can affect the absorptivity spectra of the lead chalcogenide nanocrystals ($Pb_4Te_4$, $Pb_4Se_4$, and $Pb_4S_4$). The same electric field, however, has insignificant effects on the MEG quantum probabilities and the thresholds in these nanocrystals. Furthermore, simulations show that $Pb_4Te_4$, among the aforementioned nanocrystals, has the lowest MEG threshold and the strongest absorptivity peak that is located in the multi-excitation window, irrespective of the field strength, making it the most suitable candidate for MEG applications. Simulations also demonstrate that an electric field affects the MEG characteristics in the $Pb_4Te_4$ nanocrystal, in general, less than it perturbs MEG characteristics in $Pb_4Se_4$ and $Pb_4S_4$ nanocrystals. Our results can have a great impact in designing optoelectronic devices whose performance can be significantly influenced by MEG.

# I. INTRODUCTION

Multiple exciton generation (MEG) in a semiconductor with the energy gap $E_G$ is an alternative term used for the absorption of a single photon with energy $E \geq 2E_G$ that leads to the generation of more than one exciton. Among semiconductor nanocrystals, lead chalcogenides (PbX, with X= S, Se, Te) have attracted ample attention, due to their interesting features such as small energy gaps, large bulk excitons Bohr's radii, low resistivities, large carrier mobilities, high dielectric constants, and high melting points. [1-4] These unique properties have led to the detection of MEG in the PbX nanocrystals. [5]

MEG is always in competition with other relaxation processes such as Auger recombination or phonon emission. In this respect, strong quantum confinement, phonon bottleneck [6] and relaxation of crystal momentum conservation are the main factors that accelerate MEG in nanocrystals. This conjecture was confirmed experimentally for PbX,[5,7-10] CdSe, [11-13] InAs, [14-16] Si, [17,18] and Ge [19] nanocrystals.

During the past decade, some research groups have focused on the application of the multi excitons generated in the PbX nanocrystals specifically for enhancing the efficiency of the third generation solar cells. [7,10,20,21] Moreover, solar cells share one common feature under thermal equilibrium. That is the presence of the intrinsic electric field, due to their electronic band structures. Meanwhile, the effects of $\mathscr{E}$ fields on the electronic and optical properties of semiconductor nanocrystals have been investigated intensively, in recent years. [22-27] However, none of the MEG related studies reported in the literature, so far, have considered the effect of electric fields on the MEG characteristics. Lack of such studies together with the unique properties of the PbX nanocrystals have motivated us to start this numerical investigation.

Here, we report the results of our investigations about the electronic structures of the excited states in $Pb_4X_4$ nanocrystals, using the equation of motion coupled cluster single and double (EOM-CCSD) [28,29] as a high-level ab initio approach, as implemented in the GAMESS-US program suite.[30] Although the computational method used in this work is general, we have solely considered the eight-atom $Pb_4X_4$ nanocrystals with cubic geometry. In fact, the limitations in the available computational resources and the high computational cost of the EOM-CCSD method have forced us to consider these relatively small nanocrystals and use a pseudopotential for the core electrons. To understand the physics behind our results we have also solved the problem using a two-step perturbative approach for low electric fields and compared the results with those obtained by the exact numerical solutions.

Any of the following three mechanisms can describe the MEG in a semiconductor nanocrystal:[31-33] (i) Incoherent coulomb scattering, analogous to the impact ionization mechanism in a bulk;[34,35] (ii) Coherent superposition of the single and the multiple excitons;[36-38] (iii) The direct mechanism that is based on a multi-exciton state, strongly coupled to a virtual single exciton. Hence, multiple excitons can be generated instantaneously by an absorbed photon via a perturbative process.[39,40] This mechanism is independent of the phonon coupling. Ab initio numerical methods[41-44] and the ultrafast time scale MEG experiments[18,20] have supported this explanation, and our approach is also based on this mechanism.

## II. BACKGROUND THEORY

The details of the ab initio computational method for studying the MEG process in the absence of an electric field ($\mathcal{E} = 0$), using the EOM-CCSD method are reported in Ref 32. The MEG characteristics for an unperturbed system ($\mathcal{E} = 0$) is briefly described In Appendix A. Equations (A1)-(A3) show the unperturbed MEG quantum probability, the related oscillator strength, and the optical absorptivity, respectively. An electric field alters the eigenstates, modifying the corresponding MEG quantum probability, oscillator strength and the resulting absorptivity spectrum. To evaluate these effects quantitatively, one can add the perturbing dipole moment $\left(q\mathcal{E}\cdot\mathbf{r}\right)$ directly to the Hartree-Fock Hamiltonian, $H_0$. Assuming an electric field directed in the $\hat{z}$ direction ($\mathcal{E} = \mathcal{E}\hat{\mathbf{z}}$), the total Hamiltonian becomes $H = H_0 + q\mathcal{E}z$.

Some of the states obtained for $\mathcal{E} = 0$, by the EOM-CCSD method are degenerate. To include both the degenerate and the nondegenerate states, in the perturbation approximation, we have constructed a two-step perturbation method. As also explained in Appendix B, we have used the degenerate perturbation technique to break the degeneracies, first. Then, used the first order non-degenerate perturbation theory, to calculate the new perturbed eigenfunctions and eigenenergies. The corrected MEG quantum probability obtained by the non-degenerate first order perturbation for the $k^{th}$ state ($k=1, 2, 3, \ldots, N$) is given by Eq. (B3). Comparison of Eq. (B3) with Eq. (A1) reveals that the presence of a finite electric field results in some allowed optical transitions between any given state, $k$, and the states other than $k$ can contribute in the MEG quantum probability of the $k$-state. These interactions could reduce down the transition from the single exciton to the multi exciton generation.

Moreover, an electric field also affects the dipole moment between an initial state ($\psi_i$) and a final state ($\psi_f$), modifying the related oscillator strength and hence the corresponding optical absorptivity. Employing the first order perturbation approach, the approximated closed form formulas for dipole moment, the oscillator strength, and the optical absorptivity in the presence of an electric field can be written as in Eq. (B5)-(B7).

### III. NUMERICAL SIMULATIONS

To optimize the geometry of each nanocrystal, we have used the gradient corrected DFT calculations with the B3PW91 functional. Moreover, we have employed the def2-SVPD basis set for atoms, performing the related calculations with the GAMESS-US package. The resulting maximum bond lengths, $d$, in the given cubic nanocrystals are shown in Table I. To calculate the eigenstates and the related oscillator strengths, we considered the C2V point group with $N = 200$ for each symmetry, including all valence orbitals in the active space and using the LANL2DZ basis sets with pseudopotentials representing the core electrons.

**A. The unperturbed case ($\mathscr{E} = 0$)**

The open squares (□) in Fig. 1(a)-(c) represent the MEG quantum probabilities obtained for the unperturbed $Pb_4Te_4$, $Pb_4Se_4$, and $Pb_4S_4$ nanocrystals, using the similar numerical method that we have reported in Ref 32. The solid curves represent the best fit to the scattered data, using the cubic spline approach with the smoothing parameter of 0.99. The MEG thresholds estimated from these data are given in Table I. These values correspond to the borderlines between the 'light yellow' and 'cyan' windows, shown in the figure. In the low excitation energy window (light yellow), where excitons are uncorrelated, each exciton can be accurately described within the single particle picture. The calculated data for the $Pb_4Se_4$ nanocrystal, shown in Fig 1(b), are in good agreement with the results obtained by others using the SAC-CI method.[31]

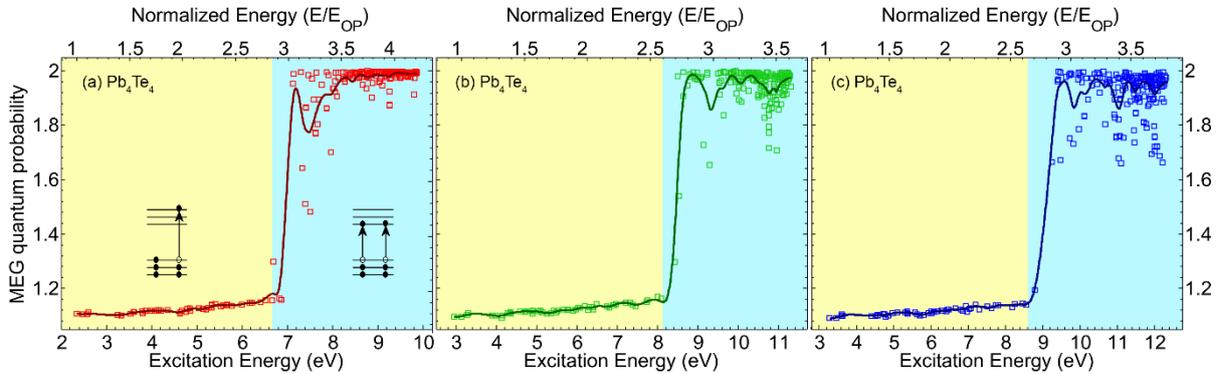

**FIG. 1.** MEG quantum probability for (a) $Pb_4Te_4$, (b) $Pb_4Se_4$, and (c) $Pb_4S_4$ nanocrystals versus energy, shown by open squares (□). The solid curves give the smoothed spline fit. The 'light yellow' and 'cyan' windows represent the single excitation and multi excitation windows.

Unlike Si,[32] Ge,[32] and CdSe,[39] all three $Pb_4X_4$ nanocrystals benefit from the similar electronic excitation behavior, having a very fast transition from the single excitation to multi excitation regime with small fluctuations in their spectra. A comparison between the estimated MEG thresholds shows that the MEG process in $Pb_4Te_4$ nanocrystal is the strongest of all, indicating that the "degree of quantum confinement" for this particular nanocrystal is the strongest among the three. We define the "degree of quantum confinement" as $\gamma \triangleq a_b/d$, as a figure of merit, where $a_b$ is the corresponding exciton Bohr's radius as shown in Table I for the given nanocrystals.[45] The corresponding "degrees of quantum confinement" are also given in the same table. As can be observed from the data given in the table, the larger the "degree of quantum confinement," the stronger the overlap between the carriers' wave functions, and hence the stronger the carriers' interactions and the smaller the MEG threshold.

Table I: The maximum bond length, $d$, exciton Bohr radius, $a_b$, MEG threshold, and the "degree of quantum confinement", $\gamma$, for unperturbed $Pb_4Te_4$, $Pb_4Se_4$, and $Pb_4S_4$ nanocrystals.

| Nanocrystal | $d$ (Å) | $a_b$ (nm)[45] | $E_{TH}$ (eV) | $\gamma$ |
|---|---|---|---|---|
| $Pb_4Te_4$ | 5.26 | 150 | 6.65 | 285.2 |
| $Pb_4Se_4$ | 4.93 | 46 | 8.1 | 93.3 |
| $Pb_4S_4$ | 4.71 | 18 | 8.6 | 38.2 |

The open circles in Fig. 2(a)-(c) represent the oscillator strengths versus the excitation energy while the solid curves show the optical absorptivity spectra corresponding to the MEG quantum probabilities, shown Fig. 1(a)-(c). Note that the main peak of the absorptivity spectrum for each nanocrystal corresponds to the largest amplitude of the corresponding oscillator strengths. One can observe that the dominant contributions to the absorptivity spectrum in the $Pb_4Te_4$ nanocrystal come from the energy states within the multi excitation window. Nonetheless, in the other two nanocrystals, the energy states within both windows contribute to the absorptivity spectra, while the contributions from the single excitation window, in either case, is greater than that of the multi excitation window. The low MEG threshold and the intense optical absorptivity in the multi excitation window for $Pb_4Te_4$ imply that MEG process in this nanocrystal is more efficient than the same processes in $Pb_4Se_4$ and $Pb_4S_4$ nanocrystals.

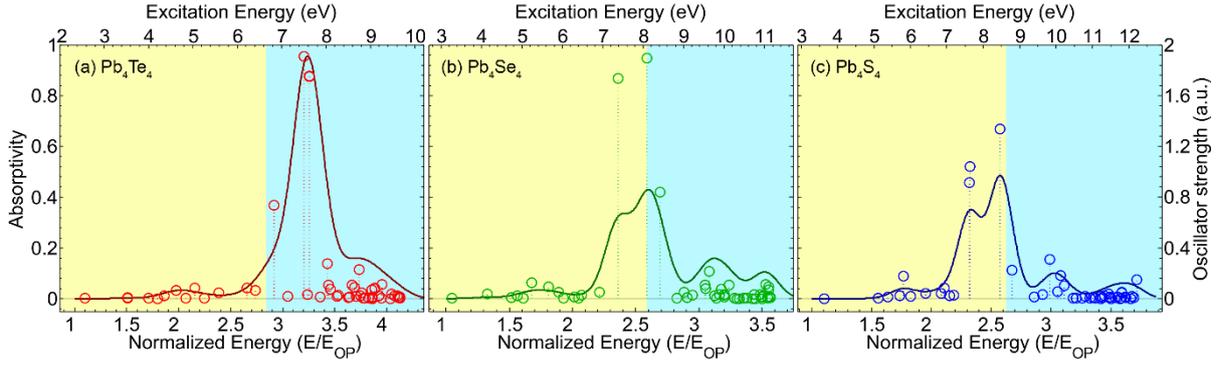

**FIG. 2.** Oscillator strengths (○) and normalized absorptivity(solid curve) for (a) $Pb_4Te_4$, (b) $Pb_4Se_4$, and (c) $Pb_4S_4$ nanocrystals versus energy. The 'light yellow' and 'cyan' windows represent the single excitation and multi excitation windows.

## B. The perturbed cases ($\mathscr{E} > 0$)

Considering various finite fields along the +z-axes ($\mathscr{E}$ = $1\times10^{-5}$, $5\times10^{-5}$, $1\times10^{-4}$, $5\times10^{-4}$, $2\times10^{-3}$, $8\times10^{-3}$, and $15\times10^{-3}$ atomic unit (au)) and using the exact numerical method, we have calculated the modified MEG quantum probabilities in $Pb_4Te_4$, $Pb_4Se_4$, and $Pb_4S_4$, first. The thin black solid curves in Fig. 3(a-c) represent the MEG quantum probability versus energy, obtained for $\mathscr{E} \leq 5\times10^{-4}$ au from the exact calculation method. The pink dots represent the exact numerical data for $\mathscr{E} = 5\times10^{-4}$ au. As it can be observed from the insets, showing the zoomed-in views around the corresponding MEG thresholds, the differences between these two sets of data are infinitesimally negligible. In other words, the exact calculations show that the electric fields $\mathscr{E} \leq 5\times10^{-4}$ have no significant effect on the MEG quantum probabilities. Moreover, dashes, dots-dashes, and the thick solid curves represent the exact numerical data for $\mathscr{E} = 2\times10^{-3}$, $8\times10^{-3}$, and $15\times10^{-3}$ au. Furthermore, we observe that the general behavior of MEG quantum probability for a given electric field differs from one nanocrystal to other. To be more specific, at any given field the MEG quantum probability in $Pb_4Te_4$ is affected the least and that in $Pb_4S_4$ is affected the most. In other words, the nanocrystal with the largest "degree of quantum confinement" is influenced the least by the electric field and vice versa.

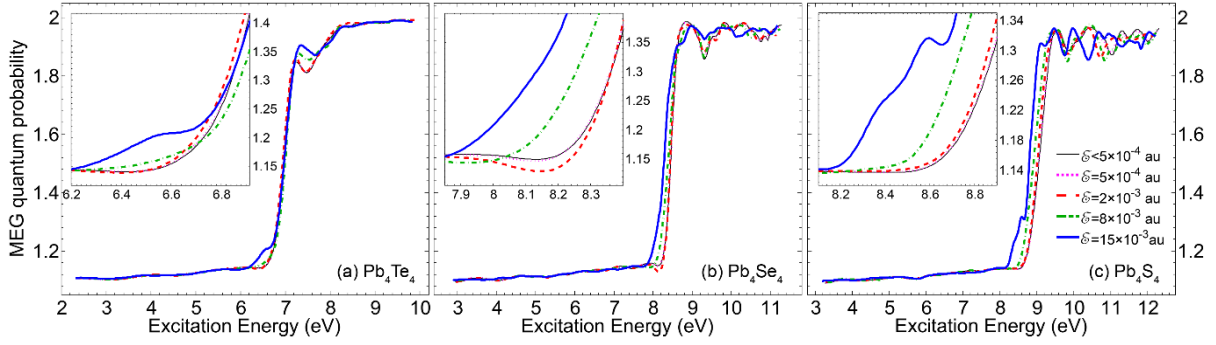

**FIG. 3.** MEG quantum probability in (a) $Pb_4Te_4$, (b) $Pb_4Se_4$, and (c) $Pb_4S_4$ for various electric fields, $0 \leq \mathcal{E} \leq 15 \times 10^{-3}$. The inset in each case illustrates a zoomed-in view for the sake of visibility.

As expected and observed from Fig. 3(a)-(c) the MEG quantum probability for each nanocrystal exhibits a specific redshift, for a given electric field. We have calculated the corresponding redshifts in HOMO-LUMO gaps by the exact numerical simulations. The resulting redshifts for $\mathcal{E} = 2 \times 10^{-3}$, $8 \times 10^{-3}$, and $15 \times 10^{-3}$ (au) are 0.5 meV, 9.8 meV, and 22.7 meV for $Pb_4Te_4$; 0.8 meV, 13.5 meV, and 46.6 meV for $Pb_4Se_4$; and 2.2 meV, 37.7 meV, and 94 meV for $Pb_4S_4$ HOMO-LUMO gaps. These results are in agreement with our expectations — i.e., the larger the electric field, the larger the redshift. However, the redshifts in HOMO-LUMO and MEG thresholds in a particular nanocrystal due to a given electric field are not correlated. Unlike the MEG threshold, the HOMO-LUMO gap is the absolute difference between two energy levels that is independent of the transitions between various states.

We have also calculated the effects of various electric fields on the oscillator strengths and the resulting absorptivity spectra. Fig. 4(a)-(c) illustrate the oscillator strengths and Fig. 5(a)-(c) show the corresponding absorptivity spectra for $Pb_4Te_4$, $Pb_4Se_4$, and $Pb_4S_4$ nanocrystals.

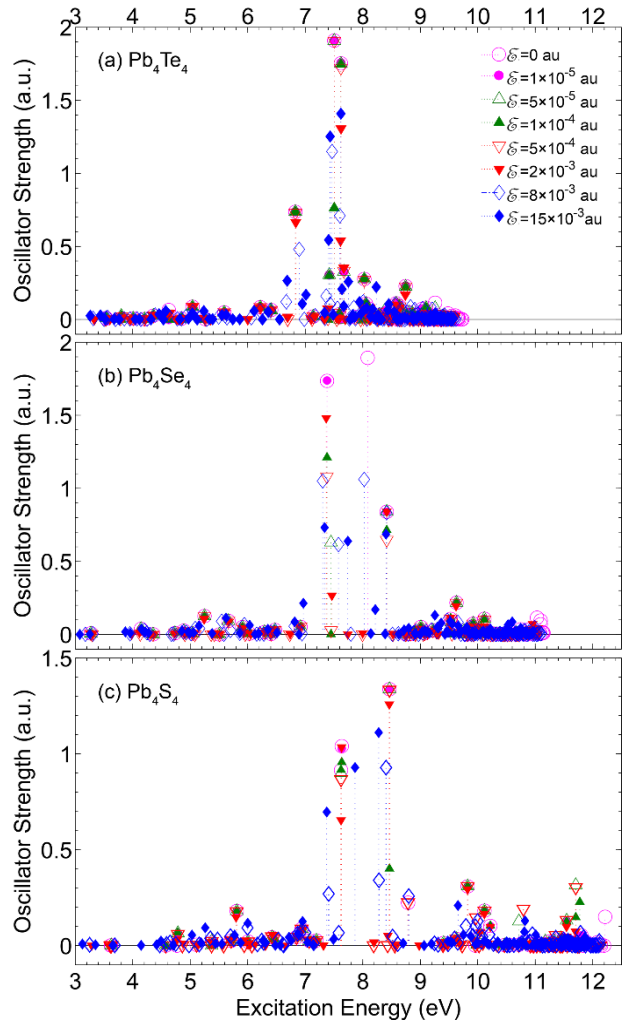

**FIG. 4.** Oscillator strengths in Pb$_4$Te$_4$ (a), Pb$_4$Se$_4$ (b), and Pb$_4$S$_4$ (c), for various electric field intensities, versus energy.

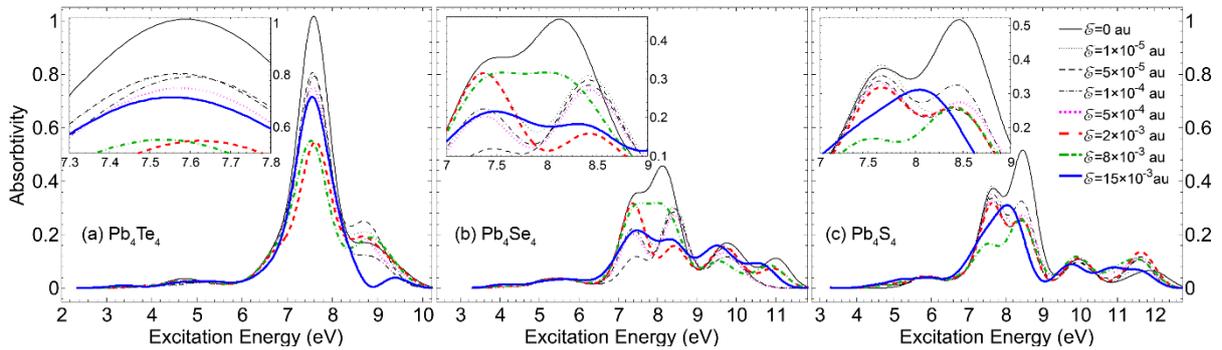

**FIG. 5.** Normalized absorptivity spectra in (a) Pb$_4$Te$_4$, (b) Pb$_4$Se$_4$, and (c) Pb$_4$S$_4$, corresponding to the oscillator strengths shown in Fig. 4. The inset in each case illustrates a zoomed-in view for the sake of visibility.

As can be seen form Fig. 4, the effect of a given electric field on the oscillator strength at a given excitation energy can be different from those related to other excitation energies. This effect depends on the degree of the state degeneracy in the absence of an electric field and the strength of the interactions between that state and the others. This can redistribute the density of states and hence the oscillator strengths. The coresponding absorptivity spectra will also be affected, acordingly, as shown in Fig. 5.

Comparison of the thin black solid curve (the unperturbed absorptivity spectrum) with other curves (the perturbed absorptivity spectra), for a particular nanocrystal, shows that any of the given electric field reduces the spectrum main peak. This can be attributed to the decrease in the overlap between the wave functions of the paired electron-hole in an exciton, due to the presence of the electric field, and hence weakening the corresponding dipole moment and the related oscillator strength. However, the trend of this reduction may not follow the field intensity as it increases. This can be seen in Fig. 5 by comparing the tiny dots, dashes, and dots-dashes representing the spectra for $\mathcal{E} = 1\times10^{-5}$, $5\times10^{-5}$, $1\times10^{-4}$ au, and those representing the absorptivities for $\mathcal{E} = 5\times10^{-4}$, $2\times10^{-3}$, $8\times10^{-3}$, and $15\times10^{-3}$ au, illustrated by thick dots, dashes, and dots-dashes, and solid curves. One may attribute this to breaking the existing degeneracies and creating new eigenstates that can redistribute the density of states and the oscillator strengths (Fig. 4) that may have a constructive or destructive effect on the absorptivity, for specific energy levels. Alterations of the main peaks of the absorptivity spectra for different field intensities observed in Fig. 5(b) and 5(c) are the manifestation of the redistributions in the density of states and oscillator strengths for $Pb_4Se_4$ (Fig. 4(b)) and $Pb_4S_4$ (Fig. 4(c)). Moreover, comparison between Fig. 3 and Fig. 5 reveals that the absorptivity, unlike the MEG quantum probability, is reduced significantly even for $\mathcal{E} = 1\times10^{-5}$ au. For example, the main peaks of the absorptivity spectra in the multi excitation windows of $Pb_4Te_4$, $Pb_4Se_4$, and $Pb_4S_4$ are reduced by 22%, 38%, and 50%, respectively. This example also verifies that while $Pb_4Te_4$ is affected the least by an electric field, $Pb_4S_4$ is affected the most.

Finally, we have evaluated the perturbed MEG characteristics, using the perturbatition technique described in Appendix B, for low electric fields ($\mathscr{E} = 1\times10^{-5}$, $5\times10^{-5}$, and $1\times10^{-4}$ au). The MEG quantum probabilities obtained by this approximate technique (Eq. B(3)) did not show any significant differences from data represented by the thin black solid curves and the tiny pink dots shown in Fig. 3 (the exact numerical data for the same field intensities). Nevertheless, the oscillator strengths and the absorptivity spectra approximated by Eq. (B6) and (B7) exhibit some deviations from the exact values shown in Fig. 4 and 5. As an example, here, we illustrate the comparison between the absorptivity spectra approximated by Eq. (B7) represented by the solid curves in Fig. 6 and the exact data for the represented by the dots in the same figure. The comparison reveals that the absorptivity spectra approximated by Eq. B(7) for $\mathscr{E} = 5\times10^{-4}$ au (red curves) match those of the exact numerical calculations (red dots), with the maximum tolerance of less than 9%, for all three nanocrystals. Moreover, the maximum tolerance in the approximated spectra for $Pb_4Te_4$ perturbed by $\mathscr{E} = 5\times10^{-5}$ au is still below 9%. However, the maximum tolerances experienced by similar spectra obtained for the other two nanocrystals, perturbed by the same field intensity, both exceed an endurable value. The deviations between approximated and exact spectra for all three nanocrystals perturbed by $\mathscr{E} = 1\times10^{-4}$ au also exceed an acceptable tolerance. Hence, the perturbation method can be confidently used for $Pb_4Te_4$ nanocrystal perturbed by electric fields of $\mathscr{E} \leq 5\times10^{-5}$ au, and for $Pb_4Se_4$ and $Pb_4S_4$ nanocrystals perturbed by $\mathscr{E} < 5\times10^{-5}$ au.

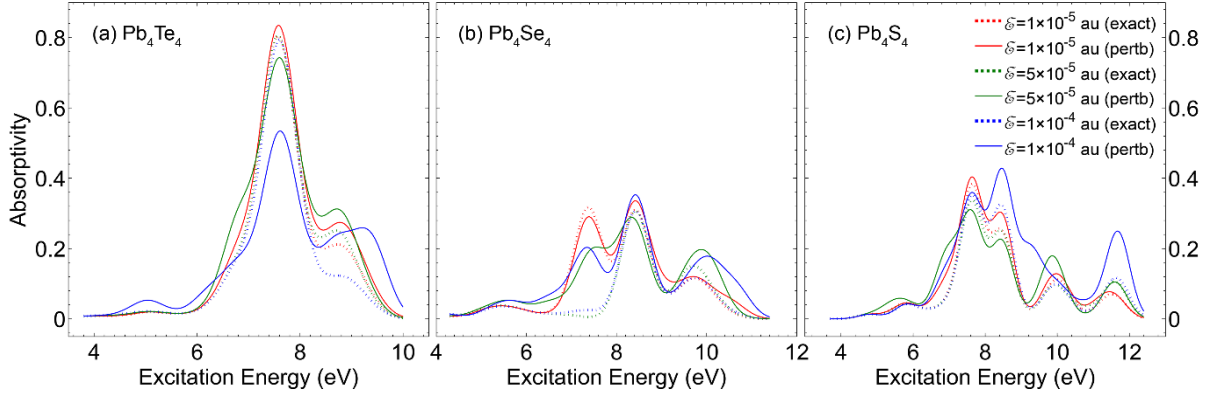

**FIG. 6.** Comparison of low field absorptivity spectra obtained by the exact (dots) and perturbation approximation (solid curves) methods.

## IV. CONCLUSION

We have analyzed the MEG processes in $Pb_4Te_4$, $Pb_4Se_4$, and $Pb_4S_4$ nanocrystals, using the EOM-CCSD framework. In this analysis, we have calculated the MEG quantum probability and spectra for the absorptivities in the absence and presence of various electric fields in the range of $1\times10^{-5}-15\times10^{-3}$ au, using the exact numerical calculations. Simulations for $\mathscr{E} = 0$, have shown that $Pb_4Te_4$, with the smallest MEG threshold and strongest absorptivity peak, is the most desirable candidate among the lead chalcogenides for MEG applications. Moreover, the main portion of the absorptivity spectrum for $Pb_4Te_4$ resides in the multi excitation window. Furthermore, we have shown that the effects of the electric fields up to $\mathscr{E} = 5\times10^{-4}$ au on the MEG quantum probabilities and the related thresholds are insignificant, for all three nanocrystals. Nonetheless, the presence of an electric field as small as $\mathscr{E} = 1\times10^{-5}$ au can have significant effect on the absorptivity spectra, for all three nanocrystals. In general, a finite electric field lowers the entire absorptivity spectrum for each nanocrystal, as compared with that for $\mathscr{E} = 0$. However, this may not follow a specific trend as the electric field is increased monotonically. Because, different field intensities may perturb various degenerate eigenstates differently, redistributing the density of states and the oscillator strengths that may have constructive or destructive effects on the absorptivity, for some energy levels. Observations also reveal that among the three lead chalcogenide nanocrystals Pb$_4$Te$_4$ with the highest "degree of quantum confinement" is the least affected one by the electric fields. The significance of the results becomes more appreciated when the goal is to design a MEG based solar cells in which the role of the intrinsic electric fields is imperative.

## ACKNOWLEDGMENTS

M. G. and M. K. M. F. would like to acknowledge the provision of the parallel computational facilities by the management and the technical staffs of the TMU IT center, for part of the numerical analysis.

H. R. gratefully acknowledges support from the UT system under the Valley STAR award.

## APPENDICES

**APPENDIX A: THE UNPERTURBED MEG CHARACTERISTICS ($\mathscr{E}=0$)**

**The MEG quantum probability**

Using EOM-CCSD method according to Ref. 32, 46 and 47 the MEG quantum probability for an unperturbed system ($\mathscr{E}=0$) is given by

$$\mathscr{P} = \frac{\sum_{n=0}^{2} n \langle \Phi | (R_{k,n})^{\dagger} R_{k,n} | \Phi \rangle}{\sum_{n=0}^{2} \langle \Phi | (R_{k,n})^{\dagger} R_{k,n} | \Phi \rangle} = \frac{\sum_{i,a}(r_a^i)^2 + 2\sum_{i<j,a<b}(r_{ab}^{ij})^2}{(r_0)^2 + \sum_{i,a}(r_a^i)^2 + \sum_{i<j,a<b}(r_{ab}^{ij})^2}, \tag{A1}$$

where $r_0$, $r_i^a$, and $r_{ij}^{ab}$ are the reference, single, and double excitation EOM amplitudes. The subscripts $i$ and $j$ represent the occupied states that can be annihilated by $\hat{i}$ and $\hat{j}$ and superscripts *a* and *b* correspond to the unoccupied states with $\hat{a}^\dagger$ and $\hat{b}^\dagger$ as the corresponding creation operators.

**The oscillator strengths and absorptivity spectrum**

The oscillator strength between two states *i* and *f*, $\mathscr{F}_{i \to f}$, is related to the unperturbed dipole moment, $\mathscr{M}_{i,f} = q \langle \psi_i^{(0)} | z | \psi_f^{(0)} \rangle$, of the corresponding transition. Under the dipole length approximation, the oscillator strength reduces to

$$\mathscr{F}_{i \to f} = \frac{8\pi^2 m_e}{3 e^2 h^2} \Delta E_{i,f} \left| \mathscr{M}_{i,f} \right|^2 = \frac{8\pi^2 m_e c}{3 e^2 h^2} \upsilon'_{i,f} \left| \mathscr{M}_{i,f} \right|^2, \tag{A2}$$

where *e* and $m_e$, *h*, and *c* are the free electron mass and charge, the Planck's constant and the speed of light in the free space, $\Delta E_{i,f} = E_f - E_i = hc\upsilon'_{i,f}$, in which $\upsilon'_{i,f}$ is the wave number related to the difference between the final and initial energy levels, $E_f$ and $E_i$.

The spectrum of the absorptivity (absorption coefficient) results from convolutions of the Gaussian functions related to the oscillator strengths:

$$\varepsilon(\upsilon') = \frac{2.175 \times 10^8}{\Delta \upsilon'} \mathscr{F}_{i \to f} \cdot \exp\left[ -2.772 \left( \frac{\upsilon' - \upsilon'_{i,f}}{\Delta \upsilon'} \right)^2 \right] \quad \text{in } \left( \text{mol}^{-1} \cdot \text{L} \cdot \text{cm}^{-1} \right) \tag{A3}$$

where $\Delta \upsilon'$ represents the full-width half-maximum (FWHM) of the Gaussian functions, in units of $\upsilon'$.

**APPENDIX B: THE PERTURBED MEG CHARACTERISTICS ($\mathscr{E} \neq 0$)**

**The corrected MEG quantum probability**

When the system is perturbed by an electric field ($\mathscr{E} \neq 0$) all degenerate states break down into the new non-degenerate eigenstates with different energies and eigenfunctions that may differ from those of the exact eigenstates. After removal all possible degeneracies by use of the degenerate perturbation technique, we use the apply the first order non-degenerate perturbation approach to obtain the corrected up wave function for the $k^{th}$ eigenstate ($k = 1, 2, 3, \ldots, N$),

$$\begin{aligned}
\left|\psi_k^{(1)}\right\rangle &= \sum_{n=1}^{N} M_{n,k} \hat{R}_n \left|\psi_0^{(0)}\right\rangle = \sum_{n=1}^{N} M_{n,k} \left( r_0(k) + \hat{R}_{S,n} + \hat{R}_{D,n} \right) \left|\psi_0^{(0)}\right\rangle \\
&= \sum_{n=1}^{N} M_{n,k} \left( r_0(k) + \sum_{i,a} r_{n,i}^{a} \left\{\hat{a}^\dagger \hat{i}\right\} + \sum_{ij,ab} r_{n,ij}^{ab} \left\{\hat{a}^\dagger \hat{i} \hat{b}^\dagger \hat{j}\right\} \right) \left|\psi_0^{(0)}\right\rangle \\
&= r_0(k) \sum_{n=1}^{N} M_{n,k} \left|\psi_0^{(0)}\right\rangle + \sum_{n=1}^{N} M_{n,k} \sum_{i,a} r_{n,i}^{a} \left\{\hat{a}^\dagger \hat{i}\right\} \left|\psi_0^{(0)}\right\rangle + \sum_{n=1}^{N} M_{n,k} \sum_{ij,ab} r_{n,ij}^{ab} \left\{\hat{a}^\dagger \hat{i} \hat{b}^\dagger \hat{j}\right\} \left|\psi_0^{(0)}\right\rangle \\
&= r_0(k) \sum_{n=1}^{N} M_{n,k} \left|\psi_0^{(0)}\right\rangle + \sum_{i,a} \sum_{n=1}^{N} M_{n,k} r_{n,i}^{a} \left\{\hat{a}^\dagger \hat{i}\right\} \left|\psi_0^{(0)}\right\rangle + \sum_{ij,ab} \sum_{n=1}^{N} M_{n,k} r_{n,ij}^{ab} \left\{\hat{a}^\dagger \hat{i} \hat{b}^\dagger \hat{j}\right\} \left|\psi_0^{(0)}\right\rangle
\end{aligned} \quad (B1)$$

where $N$ is the highest upper state used in the simulations, $\left|\psi_0^{(0)}\right\rangle$ is the unperturbed Hartree-Fock eigenfunction,

$$M_{n,k} = \begin{cases} 1 & n = k \\ qE \left\langle \psi_n^{(0)} |z| \psi_k^{(0)} \right\rangle / \left( E_k^{(0)} - E_n^{(0)} \right) & n \neq k \end{cases} \quad (B2)$$

is the dimensionless dipole moment corresponding to various $n$-states, and $\hat{R}_n = \hat{r}_0 + \hat{R}_{S,n} + \hat{R}_{D,n}$, in which $\hat{R}_{S,n}$ and $\hat{R}_{D,n}$ are the single and double excitation operators in the $n^{th}$ state.

Using Eq. (B1), one can easily obtain the modified MEG quantum probability analogues to Eq. (A1),

$$\mathscr{P}_k \simeq \frac{\sum_{i,a}\left(\sum_{n=1}^{N} M_{n,k}\ r_{n,i}^{a}\right)^2 + 2\sum_{ij,ab}\left(\sum_{n=1}^{N} M_{n,k}\ r_{n,ij}^{ab}\right)^2}{\left(\sum_{n=1}^{N} M_{n,k} r_0\right)^2 + \sum_{i,a}\left(\sum_{n=1}^{N} M_{n,k}\ r_{n,i}^{a}\right)^2 + \sum_{ij,ab}\left(\sum_{n=1}^{N} M_{n,k}\ r_{n,ij}^{ab}\right)^2}. \quad (B3)$$

As can be seen from Eq. (B2), for $n = k$ the corresponding dimensionless dipole moment becomes unity and, hence, Eq. (B3) reduces to Eq. (A1) — i.e., the MEG quantum probability for the unperturbed case. In other words, the terms representing the contributions from all the allowed transitions between the *k*-state and the *n*-states other than *k*, manifest the effect of the perturbing electric field on the MEG quantum probability of the $k^{\text{th}}$ state.

**The corrected oscillator strengths and absorptivity spectrum**

Using the corrected eigen function $|\psi_k^{(1)}\rangle$ and its Hermitian adjoint, $\langle\psi_k^{(1)}|$, one may easily obtain the perturbed dipole moment. Using the EOM-CCSD method, we have

$$\langle\psi_k^{(1)}| = \sum_{n=1}^{N} M_{k,n}^{*}\langle\psi_0^{(0)}|\hat{L}_n, \tag{B4}$$

where $\hat{L}_n$ is the de-excitation operator in the state *n*. Hence, the corrected the dipole moment up to the first order becomes,

$$\overline{\mathcal{M}}_{i,f} \simeq \langle\psi_i^{(1)}|qz|\psi_f^{(1)}\rangle \simeq q\sum_{n=1}^{N}\sum_{n'=1}^{N} M_{i,n}^{*}M_{n',f}\langle\psi_0^{(0)}|\hat{L}_n z\hat{R}_{n'}|\psi_0^{(0)}\rangle = \sum_{n=1}^{N}\sum_{n'=1}^{N} M_{i,n}^{*}M_{n'f}\mathcal{M}_{nn'} \tag{B5}$$

Using EOM-CCSD for the perturbed case the corrected oscillator strength can be written as, [45]

$$\overline{\mathcal{F}}_{i\to f} \simeq \frac{2}{3}\Delta E_{fi}\overline{\mathcal{M}}_{if}\overline{\mathcal{M}}_{if}^{\dagger} = \frac{2}{3}\Delta E_{fi}\overline{\mathcal{M}}_{if}\overline{\mathcal{M}}_{fi}^{*} \tag{B6}$$

Substituting $\overline{\mathcal{F}}_{i\to f}$ for $\mathcal{F}_{i\to f}$ in Eq. (A3), the spectrum of the perturbed absorptivity becomes,

$$\varepsilon(\upsilon') \simeq \frac{2.175\times 10^8}{\Delta\upsilon'}\overline{\mathcal{F}}_{i\to f}\cdot\exp\left[-2.772\left(\frac{\upsilon'-\upsilon'_{i,f}}{\Delta\upsilon'}\right)^2\right] \quad \text{in } \left(\text{mol}^{-1}\cdot\text{L}\cdot\text{cm}^{-1}\right). \tag{B7}$$